\documentclass[12pt]{iopart}
\usepackage{graphicx}
\usepackage{epstopdf} 

\begin{document}

\title[Magnetic Frustration on a Kagom\'e Lattice in R$_{3}$Ga$_{5}$SiO$_{14}$ Langasites]{Magnetic Frustration on a Kagom\'e Lattice in R$_{3}$Ga$_{5}$SiO$_{14}$ Langasites with R = Nd, Pr}

\author{P Bordet$^1$, I Gelard$^1$, K Marty$^1$, A Ibanez$^1$, J Robert$^2$, V Simonet$^2$, B  Canals$^2$, R Ballou$^2$, P Lejay$^3$}

\address{$^1$ Laboratoire de Cristallographie, CNRS, B.P. 166, 38042 Grenoble Cedex 9, France}
\address{$^2$ Laboratoire Louis N\'eel, CNRS, B.P. 166, 38042 Grenoble Cedex 9, France}
\address{$^3$ Centre de Recherches des Tr\`es Basses Temp\'eratures, CNRS, B.P. 166, 38 042 Grenoble Cedex 9, France}

\ead{\mailto{pierre.bordet@grenoble.cnrs.fr}, \mailto {ballou@grenoble.cnrs.fr}}

\begin{abstract}
In the R$_{3}$Ga$_{5}$SiO$_{14}$ compounds, the network R of rare
earth cations form well separated planes of corner sharing
triangles topologically equivalent to a kagom\'e lattice. Powder
samples and single crystals with R = Nd and Pr were prepared and
magnetostatic measurements were performed under magnetic field up
to 10 T in the temperature range from 1.6 K to 400 K. Analysis of
the magnetic susceptibility at the high temperatures where only
the quadrupolar term of the crystal electric field prevails,
suggests that the Nd and Pr magnetic moments can be modeled as
coplanar elliptic rotators perpendicular to the three fold axis of
the crystal structure that interact antiferromagnetically. Nonetheless, a
disordered phase that can be ascribed to geometric frustration persists
down to the lowest temperature which is about 25 times smaller than
the energy scale for the exchange interactions.

\end{abstract}


\submitto{\JPCM}

\maketitle

The langasite series, the prototype of which is the
La$_{3}$Ga$_{5}$SiO$_{14}$ compound (LGS), hence the acronym,
belongs to a vast family of materials having the
Ca$_{3}$Ga$_{2}$Ge$_{4}$O$_{14}$  type structure \cite{Mill82}.
These materials rapidly attracted a strong interest because they show
piezoelectric properties with better electro-mechanical coupling and
weaker impedance than quartz or lithium niobate and tantalate
\cite{Iwataki01}.
They can be grown as high quality large single crystals, mainly by
the Czochralski method, and are now used in surface acoustic wave
filters in telecommunication devices and high temperature sensors.
Crystallizing in a non centrosymmetric structure they exhibit quadratic
non linear optical and electrooptical properties which are also
intensively investigated \cite{Xin02}. On the other hand, probably
because the interest was focused on the striking piezoelectric
properties, the magnetic behaviors of these materials have not been
studied until now, although several compounds contain arrays of
magnetic cations. We report herein the magnetostatic properties of
Nd$_{3}$Ga$_{5}$SiO$_{14}$ (NGS) and Pr$_{3}$Ga$_{5}$SiO$_{14}$
(PGS), which are isostructural to LGS. Inspection of the crystal
structure and analysis of the magnetic susceptibility show that
these provide us with a rare opportunity for a thorough
investigation of geometric frustration in kagom\'e magnets, all
the more as large single crystals can be grown.

Geometric frustration is currently attracting strong attention for
the numerous novel phenomena it might generate \cite{Ramirez01}.
It is classically featured by macroscopic degeneracies which
prevent magnetic order to set up, thus allowing for new states of
matter and unconventional excitations to come out \cite{Lhuillier01}.
Unfortunately the materialization of these generally meets with
difficulties because of secondary interactions (next neighbors or
antisymmetric exchange and magnetoelastic interactions) or
structural and stoichiometric imperfections, which induce ordering
at low temperatures. Within the temperature range we considered,
this is not the case with NGS and PGS.

\begin{figure}[t]
\begin{center}
\includegraphics[scale=1.75]{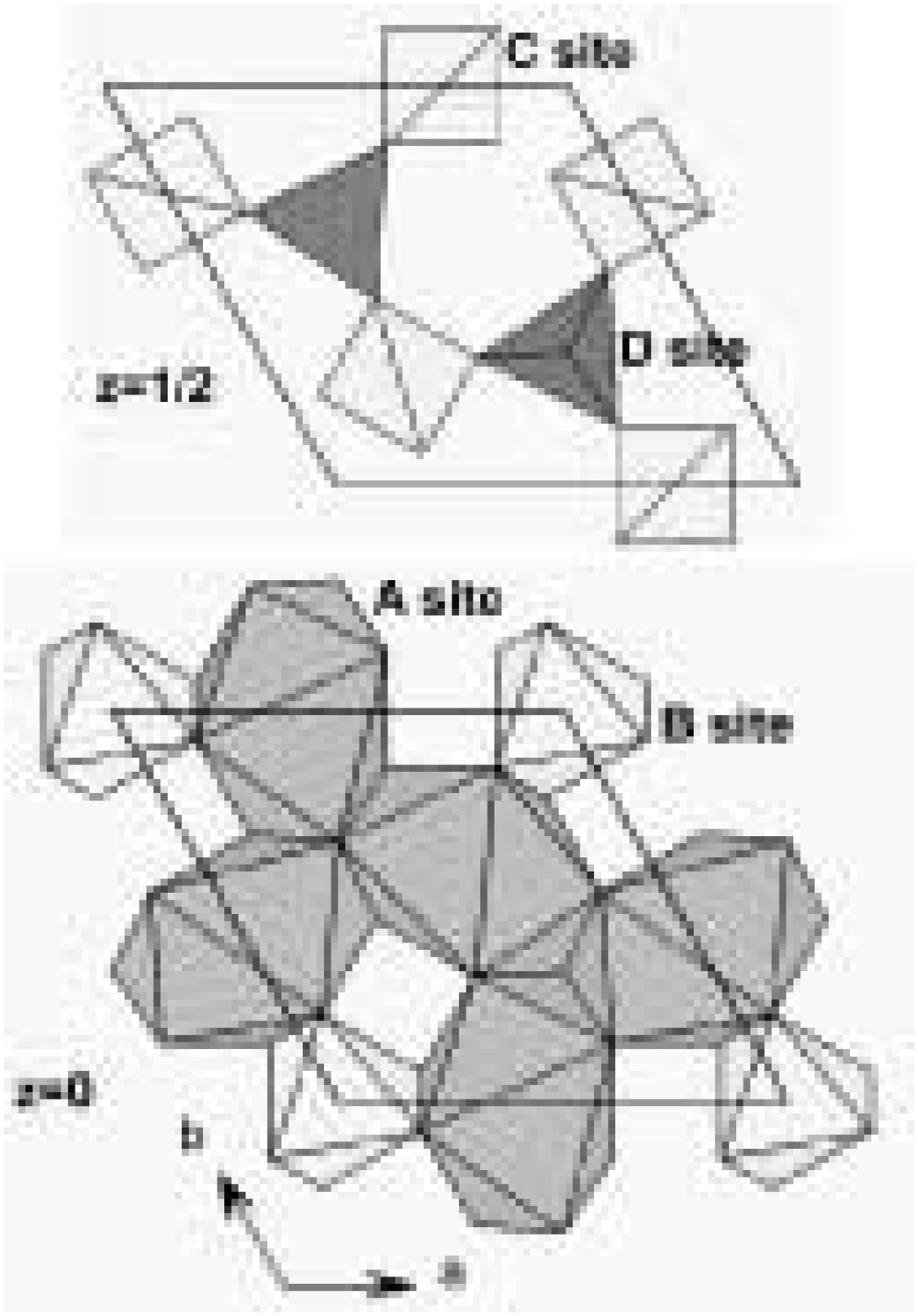}
\end{center}
\caption{Polyhedral representation of the langasite structure. Two
consecutive layers are shown. The four different cation sites are
indicated.}
\label{cryststrucvol}
\end{figure}

\begin{figure}[b]
\begin{center}
\includegraphics[scale=0.6]{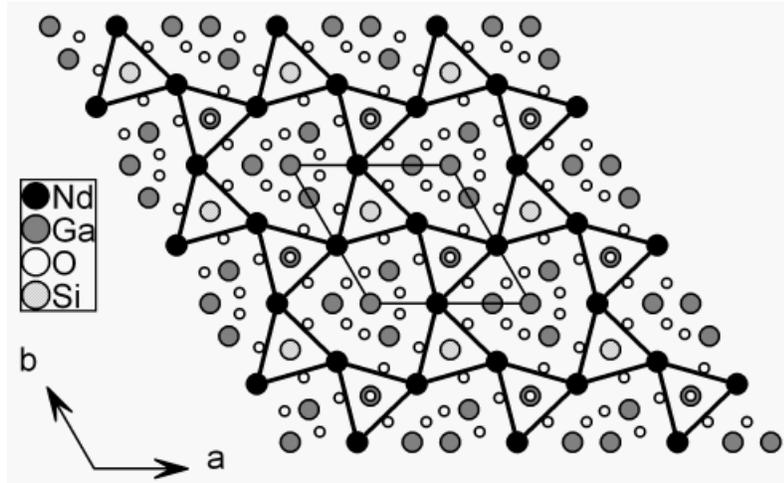}
\end{center}
\caption{Atomic arrangement of Nd$_{3}$Ga$_{5}$SiO$_{14}$
projected along the c-axis. The full lines linking the Nd$^{3+}$
cations enhance the magnetic net topologically equivalent to a
kagom\'e lattice.} \label{cryststrucpro}
\end{figure}

Langasite type compounds crystallize in the trigonal space group
P321, with lattice parameters a = b = 8.07 {\AA} and c = 5.06
{\AA} (for NGS). The structure given in Table \ref{crystdata} for
NGS and shown in Fig.\ref{cryststrucvol} contains four
crystallographically non-equivalent cation sites, and the general
formula can be written as A$_{3}$BC$_{3}$D$_{2}$O$_{14}$. The A
site (3e at ($\approx0.42$ 0 0)) is 8-coordinated by oxygen anions
forming a distorted square antiprism. The B site (1a at (0 0 0))
is octahedrally coordinated. The A and B site coordination
polyhedra share edges to form a layer at z = 0. The C site (3f at
($\approx 0.76$ 0 1/2)) and the D site (2d at (1/3 2/3 $\approx
0.53$)) are both in tetrahedral coordination, the C site being
larger than the D one. The CO$_{4}$ and DO$_{4}$ tetrahedra share
corners to form a layer centered at z = $\frac{1}{2}$, and are
also connected to the (A, B) cation layers above and below via
corner sharing. In the NGS and PGS compounds \cite{Iwataki01}, the
rare earth trivalent cations occupy the large A sites, while the
Ga$^{3+}$ cations occupy the B, C and half of the D sites. The
Si$^{4+}$ cations are localized in the remaining half of the D
sites. In Fig.\ref{cryststrucpro}, the structural arrangement of
the magnetic rare earth cations in NGS and PGS is outlined. It is
formed by planes of corner sharing triangles perpendicular to the
c-axis and separated from each other by the layer of tetrahedral C
and D sites. The triangles are equilateral by symmetry, being
centered by a 3-fold axis passing through the O1 and the mixed D
sites, and the intercationic distances are 4.18 {\AA} for NGS and
4.21 {\AA} for PGS. Exchange interactions are mediated by the O1
and O2 oxygen anions. For NGS, the Nd-O1-Nd angle is 110$^{\circ}$
with two equal Nd-O1 distances at 2.554 {\AA}, and the Nd-O2-Nd
angle is 106.5$^{\circ}$ with two Nd-O2 distances at 2.38 {\AA}
and 2.83 {\AA}. Because of the lack of hexagonal symmetry, this is
not an ideal kagom\'e lattice but considering only the shortest
atom bridging interactions we get the same overall topology.

\begin{table}[t]
\caption{Structural parameters for Nd$_{3}$Ga$_{5}$SiO$_{14}$
determined by single crystal X-ray diffraction on a Bruker-Nonius
kappaCCD diffractometer (AgK$\alpha$ radiation) on a sphere of 0.1
mm radius,  cut from a crystal grown by the floating zone
technique. Data refined using the SHELX software \cite{Sheldrick}.
R1= 0.0273, wR2 = 0.0495, GooF = 1.059. Cell parameters a = 8.066(1)
{\AA}, c = 5.062 {\AA}. The Ga3/Si3 site is half occupied by each
cation.}
\begin{center}
\begin{tabular}{|c|c|c|c|c|c|}\hline Atom & Site & X & Y & Z & U$_{\rm{eq}}$(\AA$^{2})$ \\\hline Nd & 3e & 0.41809(2) & 0 & 0 & 0.00856(2) \\\hline Ga1 & 1a & 0 & 0 & 0 & 0.01092(9) \\\hline Ga2 & 3f & 0.76479(4) & 0 & 1/2 & 0.00887(8) \\\hline Ga3/Si3 & 2d & 1/3 & 2/3 & 0.5350(2) & 0.00706(8) \\\hline O1 & 2d & 2/3 & 1/3 & 0.8042(8) & 0.0158(6) \\\hline O2 & 6g & 0.5341(4) & 0.8514(3) & 0.6916(6) & 0.0201(4) \\\hline O3 & 6g & 0.2236(4) & 0.0771(4) & 0.7610(5) & 0.0194(4) \\\hline \end{tabular}
\end{center}
\label{crystdata}
\end{table}

Powders of the NGS and PGS compounds have been prepared by solid
state reactions of stoichiometric amounts of high purity oxides at
1420$^{\circ}$C in air. X-ray powder diffraction patterns
indicated that the samples were single phase and Rietveld
refinements confirmed the crystal structures reported in Ref.
\cite{Iwataki01}. Single crystals of size up to 40 mm in length by
5 mm in diameter were grown by the floating zone method using an
image furnace, under a 99${\%}$ Ar + 1${\%}$ O$_{2}$ atmosphere,
at a growth rate of 10 mm/h \cite{Lejay}. Although it does not
allow to grow very large crystals, this technique has the
advantage of preventing a crucible pollution due to long crystal
growth times at high temperature necessary for the Czochralsky or
Brigman techniques. Moreover the size of the grown crystal is
large enough to allow probing the spatially resolved dynamic
spin-spin correlations by inelastic triple axis neutron scattering
\cite{Robert05}. The unit cell and crystallographic orientations
were checked using Laue photographs. The structure of a small
piece of a NGS crystal was re-determined using single crystal
X-ray diffraction with a kappaCCD x-ray diffractometer using
AgK$\alpha$ radiation. The results are in grood agreement with
those listed in Table \ref{crystdata} and those reported in
\cite{Iwataki01}. We shall emphasize that to the accuracy of the
experiment, the Nd site is fully occupied and no substitution of
another cation for Nd$^{3+}$ is detected.

\begin{figure}[t]
\begin{center}
\includegraphics[scale=1.25]{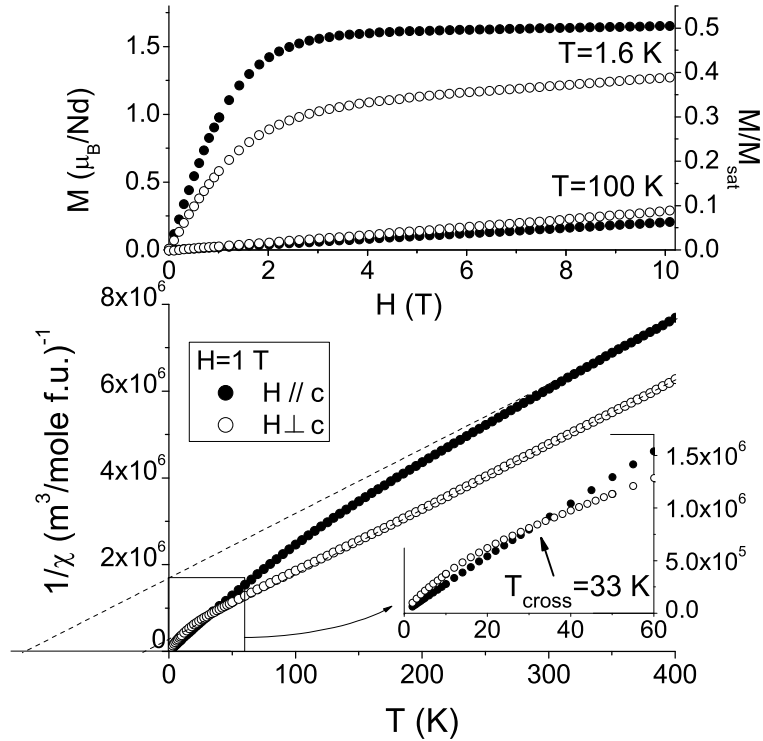}
\end{center}
\caption{Magnetostatic data measured on a NGS single crystal under
magnetic field parallel ($\|$) and perpendicular ($\bot$) to the
three fold axis $\vec{c}$. Top : Magnetic isotherms at 1.6 K and
100 K. Bottom : Thermal variation of the inverses $1/\chi_{\|}$
and $1/\chi_{\bot}$ of the magnetic susceptibilities. The dashed
lines stand for high temperature linear extrapolations for each
field direction. Inset : zoom on the cross-over of $1/\chi_{\|}$
and $1/\chi_{\bot}$.} \label{NdMagn}
\end{figure}

\begin{figure}[t]
\begin{center}
\includegraphics[scale=1.25]{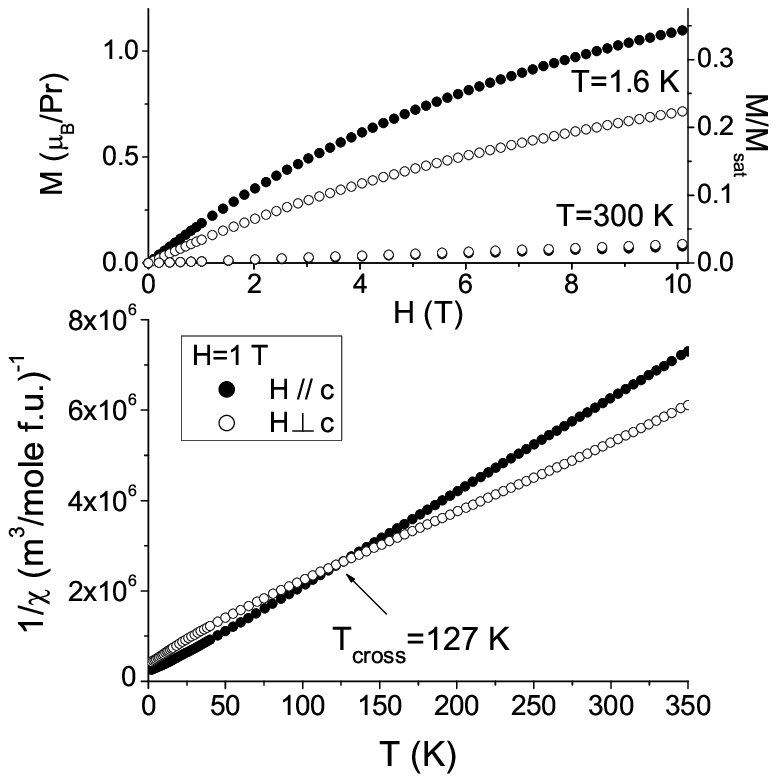}
\end{center}
\caption{Magnetostatic data measured on a PGS single crystal under
magnetic field parallel ($\|$) and perpendicular ($\bot$) to the
three fold axis $\vec{c}$. Top : Magnetic isotherms at 1.6 K and
100 K. Bottom : Thermal variation of the inverses $1/\chi_{\|}$
and $1/\chi_{\bot}$ of the magnetic susceptibilities.}
\label{PrMagn}
\end{figure}

The magnetostatic properties of NGS and PGS were investigated on
single crystals under magnetic field up to 10 T in the temperature
range from 1.6 K to 400 K on a purpose-built magnetometer using
the anti-Helmholz two-coil axial extraction method and on a
commercial Quantum Design MPMS SQUID magnetometer. We show in
Fig.\ref{NdMagn} the magnetic isotherms measured on NGS at 1.6 K
and 100 K when the magnetic field is applied parallel ($\|$) and
perpendicular ($\bot$) to the three fold axis $\vec{c}$ of the
crystal structure. We also give details of the thermal variation
of the inverses $1/\chi_{\|}$ and $1/\chi_{\bot}$ of the initial
magnetic susceptibilities deduced from the initial slope of the
magnetic isotherms measured at different temperatures and from the
thermal variation of the magnetization measured under a magnetic
field of 1 T for the same two field orientations with respect to
the $\vec{c}$ axis. Similar measurements on PGS are displayed in
Fig.\ref{PrMagn}. A large uniaxial magnetocrystalline anisotropy
is evidenced in both compounds in the whole temperature range of
the measurements with the $\vec{c}$ axis as the magnetization axis
at low temperature changing to a hard axis on increasing the
temperature at about 33 K in NGS and 127 K in PGS. A smaller
magnetocrystalline anisotropy in the plane perpendicular to the
$\vec{c}$ axis exists which however dwindles out rapidly on
increasing the temperature.

Analysis of the magnetic susceptibility can easily be performed at high
temperature (T) by limiting the expansion of the quantum statistical
average of the component of the magnetization along a quantization
axis $\vec\alpha$ (M$_{\alpha}$) to low orders in  1/T as done in
\cite{Boutron69}
\begin{equation}
\label{eq:Malpha1} M_{\alpha} = \frac{1}{N}\sum_{i = 1}^{N}
\frac{CH_{\alpha}^{i}}{T}
\{1-\frac{1}{k_{B}T}\frac{Tr\left[V^{i}O_{2\alpha}^{0}(\vec{J})\right]}{J(J+1)(2J+1)}\}
+ O(\frac{1}{T^{3}})
\end{equation}
$N$ is the number of the ions in the crystal. $C =
g_{J}^{2}\mu_{B}^{2}J(J+1)/3k_{B}$ is the Curie constant. $J$ =
9/2 for Nd$^{3+}$ ions and $J$ = 4 for Pr$^{3+}$ ions.
$H_{\alpha}^{i}$ is the component along $\vec\alpha$ of the
magnetic field on the $i$-th ion, including molecular field
contribution. $V^{i}$ accounts for the crystal electric field
potential on the $i$-th ion and can be expanded as
\begin{equation}
\label{eq:CEFH }
V^{i} = \sum_{k} \sum_{q = -k}^{k} (A_{k\alpha}^{q})^{i} O_{k\alpha}^{q}(\vec{J})
\end{equation}
in terms of Stevens equivalent operators
$O_{k\alpha}^{q}(\vec{J})$ \cite{Stevens52,Hutchings64}. $k \leq
2l = 6$ for f electrons ($l = 3$).
Using the following identity \footnote{This identity is derived on observing that
the product ${\cal{D}}_{k}$x${\cal{D}}_{l}$ of two irreducible
representations ${\cal{D}}_{k}$ and ${\cal{D}}_{l}$ of the
rotation group SO(3) contains the trivial representation
${\cal{D}}_{0}$ solely if $k = l$ and that the trace of
$O_{k\alpha}^{q}(\vec{J})O_{l\alpha}^{m}(\vec{J}$) cancels unless
$q = -m$ since under a rotation $\omega$ about the $\vec\alpha$
axis this product is multiplied by $e^{-i(q+m)\omega}$.},
\begin{eqnarray}
\label{eq:OkqO20}
Tr\left[O_{k\alpha}^{q}(\vec{J})O_{2\alpha}^{0}(\vec{J})\right] =
Tr\left[(3J_{\alpha}^{2}-(\vec{J})^{2})^{2}\right] \delta_{k,2}
\delta_{q,0} \nonumber\\
= \frac{1}{5}J(J+1)(2J+1)(2J-1)(2J+3)
\delta_{k,2} \delta_{q,0}
\end{eqnarray}
\noindent equation~(\ref{eq:Malpha1}) is considerably simplified
and allows to deduce the inverse of the initial uniform magnetic
susceptibility $\chi_{\alpha}$ along $\vec\alpha$ as
\begin{equation} \label{eq:Malpha2}
\frac{1}{\chi_{\alpha}} \approx
\frac{1}{C}\{T-\theta+\frac{(2J-1)(2J+3)}{5k_{B}}\frac{1}{N}\sum_{i
= 1}^{N} (A_{2\alpha}^{0})^{i}\}
\end{equation}
\noindent where the paramagnetic N\'eel temperature $\theta < 0$
accounts for the antiferromagnetic exchange interactions. The
$(A_{2\alpha}^{0})^{i}$ coefficients are deduced from the crystal
symmetry at the $i$-th ion position, which consists of only a two
fold axis : $\vec Z^{i} = \vec a$ for the ion at (x 0 0), $\vec
Z^{i} = \vec b$ for the ion at (0 x 0) and $\vec Z^{i} = \vec u =
\vec a + \vec b$ for the ion at (-x -x 0), where x = 0.41809(2)
for NGS (see Table \ref{crystdata}). On choosing $\vec Z^{i}$ as
the quantization axis, adopting the three fold $\vec c$ axis as
the common $\vec Y$ axis for all the ions and completing with the
appropriate $\vec X^{i}$ axis so that $(X^{i},Y,Z^{i})$ forms a
right handed frame, $V^{i}$ writes $B_{2}^{0} O_{2Z}^{0}(\vec{J})
+ B_{2}^{2} O_{2Z}^{2}(\vec{J}) + $ terms of order 4 and 6 in $k$,
where $B_{2}^{0}$ and $B_{2}^{2}$ do not depend on the $i$-th ion
since, with respect to the $(X^{i},Y,Z^{i})$ frame, the same
crystal environment is seen by the $i$-th ion. We deduce on
rotating the Stevens operators \cite{Rudowicz85} that
\begin{eqnarray}
\label{eq:A20}
\frac{1}{N}\sum_{i = 1}^{N} (A_{2\|}^{0})^{i} = - \frac{1}{2}(B_{2}^{0} + B_{2}^{2}) \\
\frac{1}{N}\sum_{i = 1}^{N} (A_{2\bot}^{0})^{i} = \frac{1}{4}(B_{2}^{0} + B_{2}^{2})
\end{eqnarray}

We observe first that $\chi_{\bot}$ does not depend upon the
orientation of $\vec\alpha$ within the plane perpendicular to
$\vec c$, which means that the in-plane anisotropy observed at low
temperature should arise from higher order terms in the crystal
electric field potential. In NGS, a linear fit of $1/\chi_{\bot}$
above 100 K and of $1/\chi_{\|}$ above 300 K using the slope
fitted from $1/\chi_{\bot}$ yields a paramagnetic N\'eel
temperature $\theta$ = -52 K, an effective moment $\mu_{\rm
eff}$ = 3.77 $\mu_B$, close to the value of the Nd$^{3+}$ free
ion, and a quadrupolar electric field parameter $(B_{2}^{0} +
B_{2}^{2})/k_{B}$ = -6.35 K. No such quantitative analysis is
possible in PGS because the temperatures at which $1/\chi_{\|}$
and $1/\chi_{\bot}$ are linear and parallel to each other are
beyond the experimental range, but negative $\theta$ and
$(B_{2}^{0} + B_{2}^{2})/k_{B}$ should be expected. As a matter of
fact $(B_{2}^{0} + B_{2}^{2}) < 0$ in both NGS and PGS since
$\chi_{\|} < \chi_{\bot}$ at high temperature.

We computed both $B_{2}^{0}$ and $B_{2}^{2}$ for NGS in a point
charge model considering the eight oxygen anions coordinating the
Nd$^{3+}$ ion and forming a distorted square antiprism around it.
Generally the as-computed absolute values differ significantly
from the actual ones, but the $B_{k}^{q}/B_{k}^{0}$ ratios are
better determined and a much greater confidence should be put on
them. We get $B_{2}^{2}/B_{2}^{0} \approx 3/2$, which suggests
that the Nd magnetic moments behave at high temperature most
probably as coplanar rotators perpendicular to $\vec c$ with an
orientational preference along the $\vec X^{i}$ axis for the
$i$-th ion within the ($\vec X^{i}$, $Y$) plane. A more complex
behavior should occur at lower temperature where higher order
terms of the crystal electric field compete with the quadrupolar
one.

\begin{figure}[t]
\begin{center}
\includegraphics[scale=1.75]{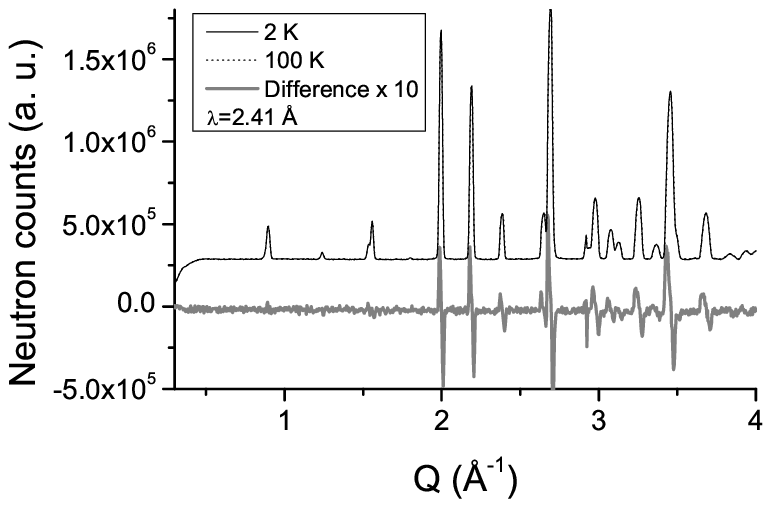}
\end{center}
\caption{Neutron patterns collected at 100 K and 2 K from a powder
sample of NGS and difference pattern multiplied by 10. The
consecutive positive and negative intensities are due to thermal
expansion.} \label{D20Exp}
\end{figure}

A fact of outmost importance was that, although finite values of
the paramagnetic N\'eel temperature $\theta$ are deduced, no
anomaly is detected in the magnetic susceptibilities of both NGS
and PGS indicating that a disordered phase would persist down to
the lowest temperature despite antiferromagnetic interactions of
one order of magnitude larger in energy scale. An elastic neutron
scattering test experiment was performed using a powder sample of
NGS on the D20 high flux diffractometer at the
Institut La\"ue Langevin \footnote{ http://www.ill.fr/YellowBook/D20}  that confirms the absence of any magnetic
ordering. As evidenced in Fig.\ref{D20Exp} no increase in the
Bragg intensity nor emergence of new Bragg intensity are observed
on the neutron pattern on decreasing the temperature from 100 K
down to 2 K.

In short, geometric frustration in Nd$_{3}$Ga$_{5}$SiO$_{14}$
(NGS) and Pr$_{3}$Ga$_{5}$SiO$_{14}$ (PGS) appears to inhibit
condensation into a N\'eel phase and to favor a disordered phase
down to the lowest temperature. We recently performed inelastic
neutron scattering experiment on both polycrystals and single
crystals evidencing liquid like dynamical spin-spin correlations
with unusual dispersive features \cite{Robert05}. As from the
present study of NGS and PGS, we expect that the series of
R$_{3}$Ga$_{5}$SiO$_{14}$ compounds with the other rare earth R
should show valuable magnetic behaviors inherent to geometric
frustration. As such they are of as much interest as from their
piezoelectric properties.

\ack{We thank the Institut La\"ue Langevin (ILL, Grenoble, France) for
providing us with beam time for the neutron scattering experiment.}

\end{document}